\documentclass[runningheads]{llncs}
\usepackage{graphicx}
\usepackage{amssymb}
\usepackage{multirow}
\usepackage{booktabs}
\begin{document}
\title{FewUser: Few-Shot Social User Geolocation via Contrastive Learning}

%
\author{
Menglin Li\inst{1}\orcidID{0000-0002-7890-7636} \and
Kwan Hui Lim\inst{1}\orcidID{0000-0002-4569-0901}
}
\authorrunning{M. Li and K. Lim.}
\institute{
Singapore University of Technology and Design, Singapore \and 
\email{menglin\_li@mymail.sutd.edu.sg, kwanhui\_lim@sutd.edu.sg}
}
\maketitle              
\begin{abstract}
To address the challenges of scarcity in geotagged data for social user geolocation, we propose \textbf{FewUser}, a novel framework for \textbf{Few}-shot social \textbf{User} geolocation.
We incorporate a contrastive learning strategy between users and locations to improve geolocation performance with no or limited training data.
FewUser features a user representation module that harnesses a pre-trained language model (PLM) and a user encoder to process and fuse diverse social media inputs effectively. 
To bridge the gap between PLM's knowledge and geographical data, we introduce a geographical prompting module with hard, soft, and semi-soft prompts, to enhance the encoding of location information.    
Contrastive learning is implemented through a contrastive loss and a matching loss, complemented by a hard negative mining strategy to refine the learning process.
We construct two datasets TwiU and FliU, containing richer metadata than existing benchmarks, to evaluate FewUser and the extensive experiments demonstrate that FewUser significantly outperforms state-of-the-art methods in both zero-shot and various few-shot settings, achieving absolute improvements of 26.95\% and \textbf{41.62\%} on TwiU and FliU, respectively, with only one training sample per class.
We further conduct a comprehensive analysis to investigate the impact of user representation on geolocation performance and the effectiveness of FewUser's components, offering valuable insights for future research in this area.

\keywords{User Geolocation \and Few-Shot Learning \and Contrastive Learning \and Social Media.}
\end{abstract}

\section{Introduction}
Leveraging users' locations has been integral for various services such as web search~\cite{kliman2015location}, recommender systems~\cite{ho2022poibert}, and targeted advertising~\cite{huang2018location}.
Such geographical information is also crucial for analyzing users' opinions on specific topics in particular regions, like COVID-19 and presidential elections~\cite{lutsai2023geolocation}. 
However, the majority of users do not provide locations in their profiles, while other location-related metadata, such as GPS tags, WiFi footprints and IP addresses, often remain inaccessible to third-party consumers~\cite{rahimi-etal-2017-neural}. 
This necessitates inferring user locations from accessible social media data, including profiles, posting content and friendship graphs. 

Social user geolocation has garnered considerable attention over the past two decades, resulting in a myriad of research efforts~\cite{do2018twitter,qiao2023twitter,li2023transformer,lutsai2023geolocation}.
Despite these advancements, the performance of existing methods significantly falls short when confronted with limited training samples, a critical issue given the prevalent data scarcity in this domain~\cite{zhou2022metageo}. 
Specifically, the activity levels of social media users vary dramatically, with some users being inactive and merely observing others' sharing without contributing any content themselves.
Additionally, geolocating in rural areas presents a typical cold-start problem.
These factors lead to a severe imbalanced distribution in social media datasets, with a small proportion of locations/labels (15\%) being associated with the majority of samples/users (80\%), resulting in a scarcity of samples for most locations. 
This underscores the critical need for advancing zero-shot or few-shot user geolocation performance.

\textbf{Contribution 1: TwiU and FliU Datasets.\footnote{Our collected datasets will be made publicly available, as permitted by the Terms of Service of the respective social media platforms.}}
Input construction and representation play a crucial role in enhancing user geolocation performance, but have not been explored in previous studies.
Existing research often adopts a uniform approach to input design~\cite{rahimi2017continuous,lourentzou2017text}, is constrained by several widely-used benchmark datasets like GeoText, which represents users as concatenated tweets with no metadata, or diverse without any justification~~\cite{ZOLA2020102312,matsumoto2022deep}, thus leading to reduced generalizability.
To address these limitations, we construct two datasets TwiU and FliU, containing rich metadata from user profiles and tweets, to enable a deep exploration of user representation on geolocation performance.


\textbf{Contribution 2: Proposed FewUser Model.\footnote{Our codes will be made publicly available after paper acceptance.}}
In this paper, we propose \textbf{FewUser}, a novel framework for \textbf{Few}-shot social \textbf{User} geolocation.
We incorporate a contrastive learning strategy between users and locations, instead of the traditional classification-based framework, to enhance the geolocation performance with no or limited training samples.
To the best of our knowledge, this is the first work to leverage contrastive learning to boost zero-shot and few-shot user geolocation capacity.
To generate efficient representations for users, we develop a user representation module that integrates diverse social media inputs and employs a user encoder to learn and merge multiple user features.
A geographical prompting module, consisting of hard prompts, soft prompts, and semi-soft prompts, is introduced to align the PLM's knowledge and geographical data and thus enhance the encoding of locations.
FewUser is trained on a contrastive loss and a matching loss jointly, with a hard negative mining strategy to refine the learning process.

\textbf{Contribution 3: Experimentation, Analysis and Findings.}
Through extensive comparative experiments with several state-of-the-art user geolocation models, we demonstrate FewUser's superiority across zero-shot and various few-shot settings.
Specifically, FewUser achieves absolute improvements of 26.95\% and 41.62\% on TwiU and FliU, respectively.
Furthermore, we perform a comprehensive analysis to investigate the impact of user representation from the perspectives of input selection, integration, and fusion, and ablation studies to explore the effects of textual backbones and geographical prompt design of FewUser, offering valuable insights for future research in social user geolocation.

\section{Related Work}

\subsection{Social User Geolocation}
Social user geolocation, the task of predicting users' locations from social media data, has garnered significant attention from both academia and industry, leading to numerous research endeavors and various shared tasks~\cite{Ramponi2023GeoLingItAE,zhang-etal-2022-changes,han-etal-2016-twitter}.
Previous works are mostly developed on several widely-used benchmark datasets, including GeoText, TwitterUS, and TwitterWorld, which represent the user as the concatenation of their tweets and contain no additional information (metadata)~\cite{huang2019hierarchical}.
The input design of these approaches is thus highly unnecessarily uniform~\cite{rahimi2017continuous,lourentzou2017text}.
While some efforts have been made to incorporate multiview fusion models by enriching input information, such as utilizing external corpora~\cite{ZOLA2020102312}, partial metadata~\cite{huang2019hierarchical}, or even weather data~\cite{matsumoto2022deep}, these studies often lack a clear justification for their input design and exhibit reduced generalizability. 
Our work addresses this gap by constructing datasets containing rich metadata information about user profiles and tweets and conducting a thorough examination of the input design issue in social user geolocation from multiple angles, as demonstrated in Section \ref{ssec:user-result}.

\textbf{User Geolocation with Contrastive Learning.}
The methods employed in the user geolocation domain have evolved from neural networks with simple structures~\cite{rahimi-etal-2017-neural,do2018twitter,thomas2018twitter} to significantly more complex ones~\cite{zhou2022metageo,liu2023ugcc,qiao2023twitter}. Furthermore, there has been a shift towards approaches based on PLMs~\cite{li2023transformer,lutsai2023geolocation}. However, all these efforts remain within the traditional classification-based framework.
Prior to our work, contrastive learning has not been incorporated into the user geolocation task, with its applications limited to related tasks such as post geolocation, as seen in efforts like ba\textit{p}tti~\cite{koudounas2023barhotti} and ContrastGeo~\cite{li2024leveraging}. 
It is important to note that social user geolocation represents a distinct and separate task from post geolocation, with the former facing more complex inputs and typically targeting a broader scope, usually at the country or global level~\cite{zheng2018survey}.

\textbf{Few-Shot User Geolocation.}
To address the data sparsity issue in user geolocation, MetaGeo~\cite{zhou2022metageo} proposes implementing few-shot user geolocation via meta-learning, which involves a complex training process~\cite{zhou2022metageo}. 
In contrast, our work presents an end-to-end user geolocation method with a straightforward training process and robust few-shot capabilities. 
Moreover, we are the first to explore open-world geolocation, that is, zero-shot geolocation, further advancing the field of social user geolocation.

\subsection{Prompting Learning}
Prompting learning involves framing tasks as "fill-in-blank" style prompts, enabling models to leverage their pre-trained knowledge for various downstream tasks~\cite{liu2023pre}. 
Initial prompting methods were primarily manual, where human experts designed task-specific prompts to guide the model's predictions~\cite{jiang-etal-2020-know}. 
These approaches showed promise in few-shot and zero-shot settings, demonstrating the flexibility of PLMs in adapting to different tasks with minimal task-specific training data~\cite{zhang2022pointclip,radford2021learning}.
As the field progressed, researchers began exploring automated methods for prompt generation, such as AutoPrompt~\cite{shin2020autoprompt} and OptiPrompt~\cite{zhong2021factual}. These methods aim to automatically discover effective prompts that can elicit the desired behavior from the model, reducing the reliance on human expertise and improving scalability.
In our work, we incorporate both manually designed prompts and automated generated prompts to provide guidance for textual backbones and bridge the modal gap between PLM's knowledge and geographical data, for better user geolocation performance with minimal training data.

\begin{figure}[!htbp]
  \centering
  \includegraphics[width=0.99\linewidth]{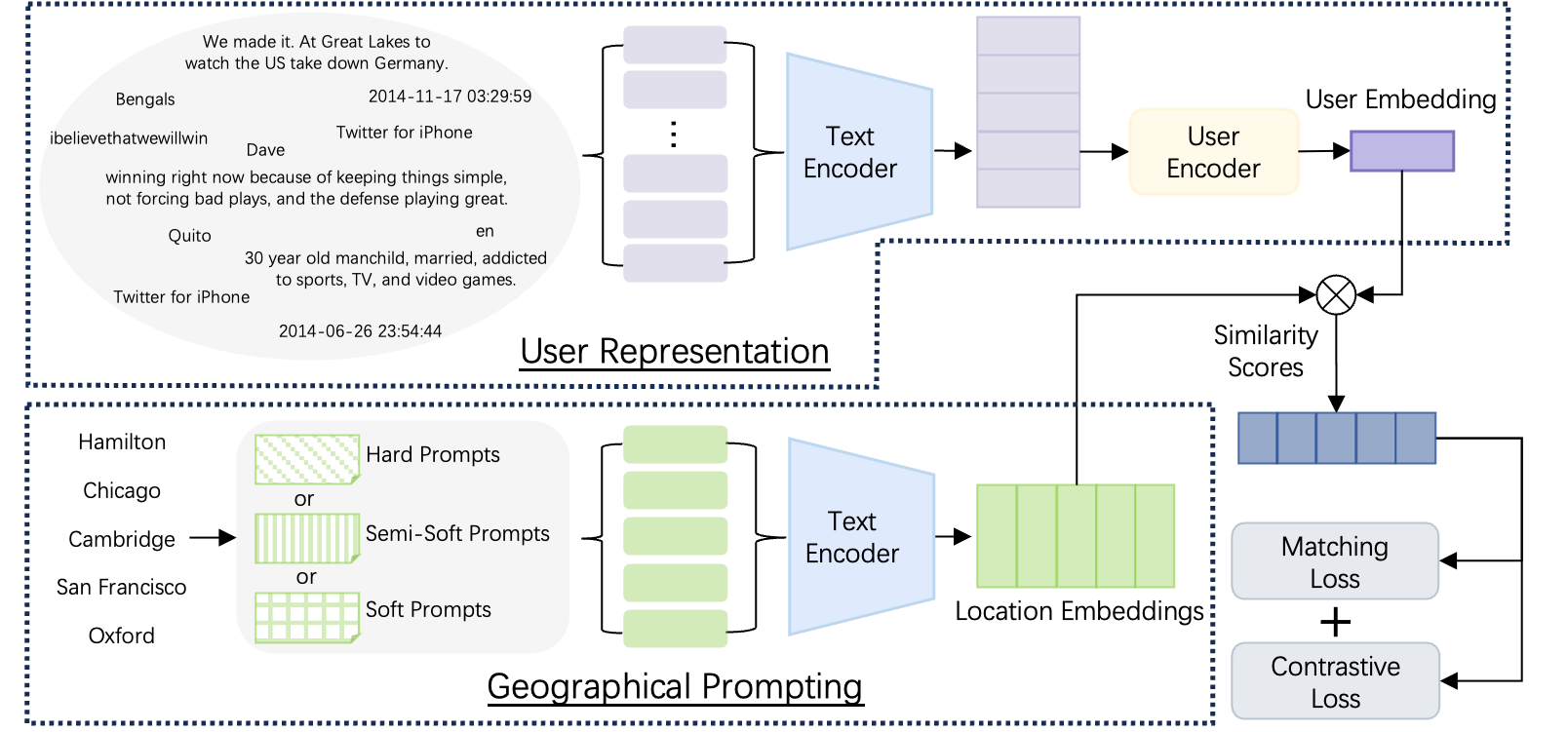}
  \caption{The Overall Framework of FewUser for Few-Shot User Geolocation.}
  \label{fig:method}
\end{figure}

\section{Our Approach: FewUser}
The overall framework of FewUser is shown in Figure~\ref{fig:method}, consisting of two shared text encoders and a user encoder.
The efficacy of FewUser hinges on effective representation learning and thus a user representation module and a geographical prompting module are incorporated to generate efficient representations for users and locations, respectively.
Contrastive learning is then implemented via two training objectives, a contrastive loss and a matching loss, with a hard negative mining strategy to refine the learning process.




\subsection{User Representation}\label{ssec:input}

\begin{figure}[!htbp]
  \centering
  \includegraphics[width=0.99\linewidth]{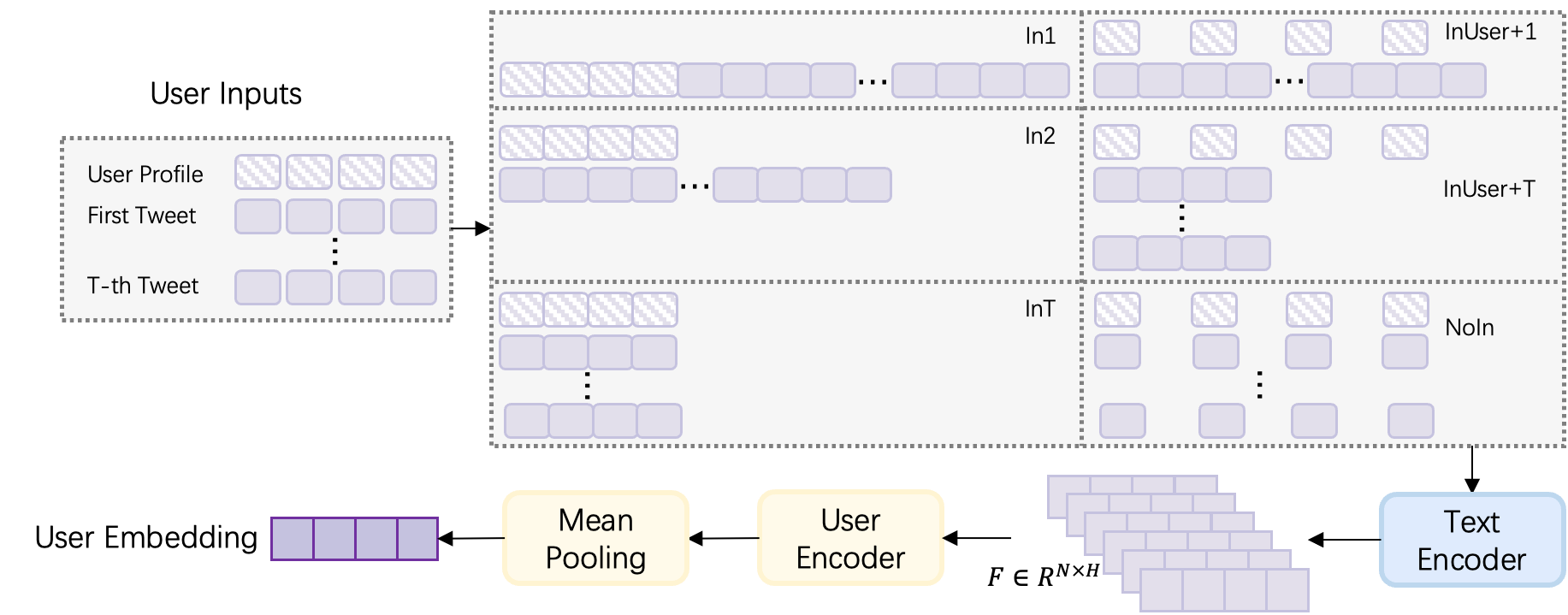}
  \caption{User Representation Module.}
  \label{fig:input}
\end{figure}
This module is introduced to integrate diverse social media inputs, encode them using the text encoder, and fuse multiple features into a comprehensive user representation via a user encoder.

\textbf{Integration.}
We design strategies to integrate diverse inputs and incorporate various structures of user encoders to fuse multiple user features, as displayed in Figure~\ref{fig:input}.
The available inputs encompass user profiles (user names, locations, descriptions, etc) and historical tweets (tweet text, source, hashtags, timestamp, and so on).
As a result, the number of input fields may exceed thirty, even with a limited number of tweets, like six tweets.
Consistent with prior studies, we primarily focus on textual information for estimating user locations and most inputs are short texts of free form.
Thus we propose six approaches to integrate these inputs to enhance computational efficiency and potentially improve model performance.
Assuming the number of tweets is $T$, the six strategies are:
\begin{itemize}
    \item \textbf{In1}: Merge all input columns into a single sentence.
    \item \textbf{In2}: Combine user profile-related columns into one sentence and tweet-related fields into another, resulting in two sentences.
    \item \textbf{InT}: Aggregate user profiles-related columns into one, and group columns related to each tweet into individual sentences, yielding $T+1$ sentences.
    \item \textbf{InUser+1}: Integrate tweet-related columns into a single sentence, while keeping user-related columns separate.
    \item \textbf{InUser+T}: Retain the original user-related columns unchanged and combine columns related to each tweet into separate sentences.
    \item \textbf{NoIn}: No concatenation.
\end{itemize}

Then we get $N$ input sentences after integration and use a PLM as the text encoder to perform encoding.
Coupled with [CLS] token extraction operation, a feature matrix $F \in \mathbb{R} ^{N\times H}$ is obtained, where $H$ is the text encoder's hidden size.
For subsequent contrastive learning between users and location, the obtained user features need to be transformed into one.

\textbf{Fusion.}
We further employ a user encoder on $F$ to learn feature-wise correlations and apply mean pooling to keep dimensional consistency of representations, resulting in the user embedding $u \in \mathbb{R}^{H}$.
Three types of user encoders are taken into consideration.
The first type is time series models, such as LSTM, BiLSTM, RNN, and GRU, to better capture temporal information (timestamp) involved in inferring locations.
The second one is time-insensitive but with prevalent applications across various domains, including transformer encoder, bottle-neck adapter, and multi-layer perception.
The last one is the simplest method and parameter-free, mean pooling.

In addition to integration and fusion, we investigate input selection as well and perform an analysis of how to generate efficient representations as shown in Section \ref{ssec:user-result}.

\subsection{Geographical Prompting}

To bridge the modal gap, we develop a geographical prompting module, containing hard, soft, and semi-soft prompts, to activate the text encoder for better encoding of geographical data.


\textbf{Hard Prompts.}
We utilize the task description of social user geolocation as the original prompt.
To enhance lexical diversity while maintaining fidelity to the original prompt, we generate prompts through paraphrasing, producing semantically similar or identical expressions.
For instance, the original prompt "\textit{A user resides in [CLASS].}" may be paraphrased to "\textit{A user from [CLASS].}".
Location names are inserted into the class token slot to apply the prompt. 
Additionally, we incorporate Question-Answering (QA) prompts to further increase diversity.
In this case, the same original prompt is converted into "\textit{Question: Where does this user reside in? Answer: [CLASS].}"
We refer to these prompts as hard prompts, as they remain fixed during the training process.

\textbf{Soft Prompts and Semi-Soft Prompts.}
While hard prompts are effective, their usage is constrained by the manual effort and guesswork required to craft suitable prompts. 
Furthermore, they provide a lower-bound estimate of the model's performance in terms of factual information encoding. 
Inspired by OptiPrompt~\cite{zhong2021factual}, we refine this estimate by directly optimizing the representation of prompts in a continuous embedding space. 
Specifically, the prompt is defined as \[t_{prompt} = [V]_1  \,\, [V]_2 \,\, \cdots \,\, [V]_m \,\, [CLASS], \] where each $[V]_{i} \in \mathbb{R}^{H}$ is a dense vector with the same dimension as the text encoder's hidden size and $m$ is the number of $[V]$ tokens, a predefined hyperparameter.
These dense vectors are learnable during training by adding new tokens to the text encoder's vocabulary. 
There are two approaches to initializing the $[V]$ tokens: random initialization, referred to as soft prompt, and a more sophisticated method that uses hard prompts to determine the number and layout of $[V]$ tokens, initializing each $[V]_i$ with the input embedding for the corresponding tokens in hard prompts.
The latter approach is named semi-soft prompt.

\subsection{Training Objectives}\label{ssec:loss}
Contrastive learning is implemented through a contrastive loss and a matching loss, inspired by ALBEF~\cite{li2021align}, based on user and location representations.

\textbf{Contrastive Loss} is to minimize the distance between positive pairs and maximize the distance between negative pairs.
Consider a user $u$, its embedding is obtained via the user representation module, denoted as \(g_u(u)\).
For each dataset, we use the names of all the classes in the dataset as the set of potential location pairings, represented as \(\{l_j\}_{j=1, ..., K}\).
The transformation \(g_l(l_j)\) converts location names into sentences and generates location representations.
A pair is deemed positive if the location corresponds to the ground-truth label of the user, denoted as \(l^+\).
Utilizing the dot product as a measure of similarity, we adopt InfoNCE~\cite{he2020momentum}, a form of contrastive loss function:
\begin{equation}\label{eq:contrast}
    \mathcal{L}_{contrast} = -\log \frac{exp(g_u(u) \cdot g_l (l^+)/\tau)}{\sum_{j=1}^{K}exp(g_u (u)\cdot g_l (l_j)/\tau)},
\end{equation}
where \(\tau\) is a temperature hyperparameter.
This loss can be interpreted as the log loss of a $K$-way softmax-based classifier that aims to classify $u$ as $l^+$.

\textbf{Matching Loss} is to determine whether a user-location pair is positive (matched) or negative (unmatched). 
To achieve this, We employ the fused representation of the user-location pair, and attach a fully-connected (FC) layer followed by a softmax function, to compute a probability distribution $\mathbf{p}$ over $(k+1)$ classes probability $\mathbf{p}$, where $k$ represents the count of hard negatives. 
The matching loss is defined as:
\begin{equation}\label{eq:match}
    \mathcal{L}_{match} = \mathbb{E}_{(u, l) \sim \mathcal{D}} [F(\mathbf{y}(u, l), \mathbf{p}(t, l))],
\end{equation}
where $\mathbf{y}$ is a $(k+1)$-dimensional one-hot vector representing the ground-truth label.

To generate the joint representation, we explore three fusion designs—Cross-Attention (CA), Sum, and Concat types—to fuse user and location representations. 
The Concat type, which applies feature-wise concatenation, and appends a bottle-neck adapter as the fusion encoder followed by an MLP layer to keep dimension consistent, proves to be the most effective. 

For mining $k$ hard negatives from the $K-1$ locations (excluding the ground-truth location $l^+$), we present two approaches: multinominal and top.
The first one samples negative locations based on a multinominal distribution weighted by user-location similarity, while the second approach selects the top-$k$ locations with the highest similarity scores, providing the most challenging negative examples for robust model training.

The full training objective of FewUser is:
\begin{equation}
    \mathcal{L} = \mathcal{L}_{contrast} + \mathcal{L}_{match}. 
\end{equation}

\section{Experiment Results}
\subsection{Experiment Setting}
\textbf{Datasets.}
We construct two datasets, TwiU and FliU, containing richer information as compared to existing benchmarks like TwitterWorld and GeoText~\cite{huang2019hierarchical}.

TwiU builds upon the WNUT16 dataset~\cite{han-etal-2016-twitter}, which we further enhance by querying Twitter's API to retrieve user profiles, text and metadata of tweets using tweet IDs.
WNUT16 provides label information, including the name, latitude, and longitude of the user city.
This dataset consists of 7,789 users from 1,261 cities worldwide, with information on user names, user descriptions, user creation date-times, user locations, user languages, user time zones, tweet texts, tweet creation date-times, posting sources, and hashtags.

FliU is derived from the YFCC-100M dataset~\cite{thomee2016yfcc100m}.
We extract geotagged Flickr posts with a minimum accuracy level of city granularity.
The geographical scope is restricted to the United States (US) using boundary information from GeoNames\footnote{https://download.geonames.org/} and user time zones.
The dataset is further enriched by crawling user profiles via Flickr's API.
Labels for the user geolocation task are generated by matching and standardizing the city field in user profiles using the US city list on GeoNames.
The FliU dataset comprises 11,395 users from 1,688 distinct US cities, including user-related fields like user names, profile descriptions, occupations, hometowns, countries, user join date-times, and post-related columns such as titles, descriptions, user tags, machine tags, devices, (photo) take date-times, and upload date-times.

To adapt these datasets for the few-shot user geolocation task, we perform several preprocessing steps.
Firstly, part of minority classes are excluded due to insufficient samples.
Each dataset is then split into training, development, and test sets with a category-by-category random ratio of 0.7:0.15:0.15.
The development set is downsampled to rebalance the original imbalanced distribution for few-shot training, while the test set remains unaltered to ensure fair evaluation.
To alleviate the influence of randomness in training sample selection, we create three $s$-shot training subsets from each dataset's training set using different random seeds, with $s$ ranging from 1 to 8.
The reported experimental results are the average performance of the model across three training subsets.

\textbf{Evaluation Metrics.} 
In assessing the performance of user geolocation, we utilize three metrics: accuracy ($acc$), mean distance error ($meanD$), and median distance error ($medD$). 
Given that the datasets span either country-wide or global scales, we express the latter two metrics in kilometers. 
It is important to note that only the TwiU dataset contains the coordinate information of labels, thus the calculation of error distances is confined to this dataset.

\textbf{Training Details.}
We implement FewUser and baselines using the \textit{transformers} package.
For the text encoder, we employ SimCSE (sup-simcse-bert-large-uncased)~\cite{gao-etal-2021-simcse} as the backbone, and use the [CLS] token as the sentence embedding.
Model parameters are optimized with AdamW optimizer, and the early-stopping mechanism is applied.
In the 8-shot setting on TwiU, we conduct a hyperparameter search and select a training batch size of 8, an evaluation batch size of 1, 100 epochs, a learning rate of 8e-6, a temperature $\tau$ of 0.03, and AdamW optimizer setting with beta1 and beta2 values of 0.85 and 0.999, respectively.

FewUser is trained using a dual-objective strategy, incorporating both contrastive and matching losses.
For user representation, we consider user profiles, tweet text, and metadata of tweets, from the latest six tweets of each user as inputs, which are integrated using In1, and use the text encoder to generate user embedding.
Location embedding is obtained via a semi-soft prompt, initialized from \textit{"I'm in [CLASS]."}.
We sample six hard negatives from a multinomial distribution weighted by similarity scores to compute the matching loss.
Note that unless specified otherwise, all experimental settings will adhere to the configurations described in this section.

\textbf{Inference.}
During inference, we utilize contrastive similarity scores between the target user and all unique prompted locations in the dataset to identify the most probable (user, location) pair, thereby determining the predicted location according to FewUser.
Unlike the training process, there is no calculation of the contrastive loss and matching loss during the inference phase.

\textbf{Baselines.}
We benchmark FewUser against several established social user geolocation models, including HLPNN~\cite{huang2019hierarchical}, MENET~\cite{do2018twitter}, GeoNN~\cite{lourentzou2017text}, and GeoDare~\cite{rahimi-etal-2017-neural}. 
Additionally, we include two post geolocation approaches, transTagger~\cite{li2023transformer} and GeoBERT~\cite{scherrer2021social}, as they both utilize PLMs like BERT for location estimation, which are closely related to our work.
Furthermore, we introduce a variant of FewUser, named ClassUser, which replaces the contrastive learning component with a traditional cross-entropy loss for location classification while retaining the rest of the framework.

\begin{table*}[!htb]
  \caption{Performance comparison between FewUser against geolocation models under 0, 1, and 8-shot settings, on TwiU and FliU in terms of $acc$ (\%), $meadD$ (km), and $medD$ (km).}
  \label{tab:main}
  \centering
  \resizebox{1\textwidth}{!}{
  \begin{tabular}{l|ccccccccc|ccc}
    \toprule
        & \multicolumn{9}{|c|}{TwiU} & \multicolumn{3}{c}{FliU} \\ 
          \cline{2-13}
        & \multicolumn{3}{|c}{0-Shot} & \multicolumn{3}{c}{1-Shot} & \multicolumn{3}{c|}{8-Shot} & 0-Shot & 1-Shot & 8-Shot\\ 
          \cline{2-13}
Model   & $acc$ & $meanD$ & $medD$ & $acc$ & $meanD$ & $medD$ & $acc$ & $meanD$ & $medD$ & $acc$ & $acc$ & $acc$ \\
    \midrule
GeoBERT&	3.83& 	7602 &	6772 &	10.21 &	5963 &	4689 &	31.10 &	2086 &	353 	&5.23 	&7.08 &	10.57 \\
GeoDare&	7.66& 	7489 &	8335 &	10.69 &	6516 &	6637 &	18.50 	&5245 	&1119 	&3.49 &	7.34 &	7.40 \\
GeoNN&	4.31& 	11402 &	10278 &	11.00 &	7809 &	7716 &	22.97 &	4539 &	917 &	4.75 &	4.38 	&8.93 \\
MENET&	4.31 &	8777 &	10646 &	13.56 &	5760 &	2132 &		17.38 &	4131 	&1039 &	8.03 &	9.46 &	8.24 \\
HLPNN&	11.48 &	9175 &	11716 &	25.36 &	4330 &	776 &36.20 &	3401 &	375 &	3.96 &	9.35 &	3.86 \\
transTagger&	2.87 &	9622 &	9499 &	17.54 &	5347 &	3422 &	33.49 &	2746 &	373 &	9.19 &	8.45 	&28.79 \\
\hline
ClassUser&	3.83 &	10671 &	14124 &	16.11 &	4302 &	1092 &	60.13 &	918 	&34 	&6.18 	&11.46 &	43.95 \\
FewUser&	\textbf{16.75} &	\textbf{4533} &	\textbf{906} &	\textbf{52.31} &	\textbf{1680} &	\textbf{5} &	\textbf{71.77} 	&\textbf{768} &	\textbf{0} 	&\textbf{27.73} &	\textbf{51.08} &	\textbf{56.15} \\
  \bottomrule
  \end{tabular}
  }
\end{table*}

\subsection{Zero-Shot User Geolocation}
We evaluated the zero-shot geolocation performance of FewUser on the TwiU and FliU datasets, and observed that FewUser out-performs all baselines by several multiples, as shown in Table \ref{tab:main}. 
Without any training samples, FewUser achieves a notable accuracy of 27.73\% on FliU. 
For TwiU, FewUser attains a slightly lower performance of 16.75\%, attributed to the absence of specific downstream adaptations.
Compared with other models based on traditional classification approaches, our method demonstrates significant advantages in zero-shot user geolocation. 
The strong performance of FewUser underscores the potential of using contrastive learning for zero-shot geolocation tasks.

\begin{figure}[!htb]
  \centering
  \includegraphics[width=0.99\linewidth]{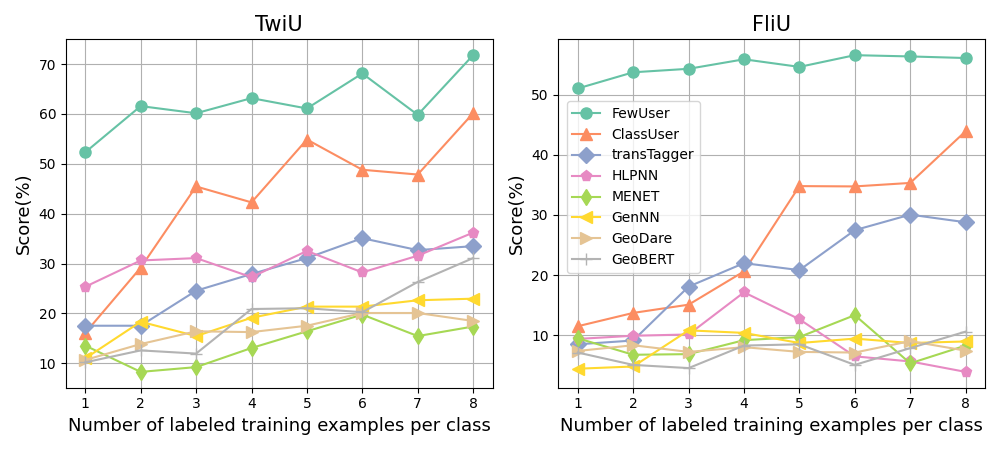}
  \caption{Few-shot performance comparison between FewUser, ClassUser, and representative geolocation models, on TwiU and FliU in terms of accuracy.}
  \label{fig:main}
\end{figure}

\subsection{Few-Shot User Geolocation}
Our evaluation of FewUser encompasses various few-shot settings, including 1, 2, 3, 4, 5, 6, 7, and 8 shots.
Figure \ref{fig:main} depicts the few-shot performance of our proposed model and baselines in terms of accuracy.
By introducing user representation, geographical prompting, and dual training objectives, FewUser demonstrates powerful geolocation capacity across different few-shot scenarios.

The results indicate that FewUser consistently surpasses comparative models by a substantial margin. 
Notably, FewUser excels particularly when limited samples per category are available. 
For example, in a 1-shot scenario with only one training sample, FewUser achieves absolute improvements of 26.95\% and 41.62\% on TwiU and FliU, respectively, compared to the competitive model HLPNN.
In addition to accuracy, the superiority of FewUser is also evident in the mean and median error distances, as shown in Table \ref{tab:main}.
Particularly noteworthy is the results from zero-shot to 1-shot, adding just one training sample per category enables FewUser to improve its accuracy by 35.56\% on TwiU.

The baselines exhibit poorer performance on FliU compared to TwiU. 
This may be attributed to the fact that these models are developed based on Twitter datasets and exhibit limited generalizability to the Flickr platform. 
Moreover, as a photo-sharing platform, Flickr contains less rich information compared to Twitter. 
In contrast, our model demonstrates strong generalizability across both social media platforms. 

Furthermore, the results reveal that FewUser shows significant improvement over the variant ClassUser under all zero-shot and few-shot settings, indicating the efficacy of incorporating contrastive learning into user geolocation. 
Additionally, ClassUser outperforms baselines in most few-shot settings, suggesting that besides the contrastive learning component, the remaining elements of our framework design are also effective for few-shot social geolocation.

\subsection{User Representation}\label{ssec:user-result}
In this section, we conduct comprehensive experiments with FewUser and ClassUser on the TwiU and FliU datasets, to examine the impact of user representation on the performance of few-shot social user geolocation, for both contrastive learning-based and classification-based models across different social platforms. Specifically, we perform these experiments under an 8-shot setting.


\begin{table*}[!htb]
  \caption{Performance of FewUser and ClassUser on TwiU and FliU with the number of tweets ranging from 2 to 10.}
  \label{tab:tweets}
  \centering
  \begin{tabular}{l|l|ccccccccc}
    \toprule
    & & \multicolumn{9}{c}{Number of Tweets}  \\ 
    \cline{3-11}
     & Model    & 2 & 3 & 4 &5 &6 &7 &8 &9 &10\\
    \midrule
    \multirow{2}{*}{TwiU} & FewUser	&65.55 	&66.51 	&66.83 	&67.78 	&\textbf{71.77} 	&67.46 	&67.30 	&67.15 	&70.18\\
    & ClassUser &56.94 	&59.17 	&57.10 	&56.94 	&\textbf{60.13} 	&57.89 	&55.98 	&55.34 	&58.37 \\
    \hline
    \multirow{2}{*}{FliU} & FewUser	&54.89 	&55.78 	&54.46 	&55.26 	&56.15 	&55.78 	&56.26 	&56.58 	&\textbf{56.95} \\
    & ClassUser &42.79 	&42.26 	&43.58 	&\textbf{43.95} 	&\textbf{43.95} 	&43.26 	&43.58 	&39.57 	&42.16 \\
  \bottomrule
  \end{tabular}
\end{table*}


\textbf{What information is useful?} We investigate this problem from: the number of tweets and field selection. 
Older tweets may not contribute to accurate predictions of a user's current location, and using an excessive number of tweets could significantly increase computational overhead. 
As shown in Table \ref{tab:tweets}, we vary the number of tweets from 2 to 10 and find that using six tweets consistently yields the best performance. 
While information is abundant, not all of it is beneficial for geolocation.
For instance, the posting time of tweets might hinder model learning, particularly for models not specifically designed for time-series data. 
We thus compare models using all inputs (All) with those excluding posting time (NoPostTime) or all tweet metadata (NoPostMeta). 
The results in Table \ref{tab:column} suggest that for classification-based models like ClassUser, NoPostTime is a favorable choice. 
Conversely, for contrastive learning-based models like FewUser, using all available inputs tends to perform better on Twitter.

\begin{table*}[!htb]
  \caption{Performance of FewUser and ClassUser on TwiU and FliU with different input sets.}
  \label{tab:column}
  \centering
  \begin{tabular}{l|ccc|ccc}
    \toprule
    & \multicolumn{3}{c|}{TwiU} & \multicolumn{3}{c}{FliU}  \\ 
    \cline{2-7}
    & All & NoPostTime & NoPostMeta & All & NoPostTime & NoPostMeta \\
    \midrule
FewUser&	\textbf{71.77}& 	68.26& 	64.91& 	56.15& 	\textbf{56.42}& 	50.03 \\
ClassUser&	60.13& 	\textbf{62.84}& 	56.62& 	43.95& 	\textbf{44.69}& 	37.30 \\ 
  \bottomrule
  \end{tabular}
\end{table*}

\textbf{How to integrate inputs?} To determine the optimal method for effectively and efficiently integrating inputs, we experiment with six integration approaches as proposed in Section \ref{ssec:input}, with the results presented in Table \ref{tab:integrate}. 
Surprisingly, for FewUser, the simplest method, In1, yields the best performance. 
This may be attributed to the fact that contrastive learning demands a higher degree of alignment in the embedding space of paired entities. 
In1 ensures that user and location representations reside in the same embedding space, as it does not involve post-processing for user features.
For ClassUser, In2 emerges as the superior approach, further supporting the aforementioned hypothesis. 
The results also indicate that an excessive number of input features can be detrimental to performance. 
Moreover, the computational burden is significantly increased, particularly when using PLM as the text encoder.
\begin{table*}[!htb]
  \caption{Performance of FewUser and ClassUser on TwiU and FliU on the integration type. InU+1 and InU+T are short for InUser+1 and InUser+T.}
  \label{tab:integrate}
  \centering
  \resizebox{1\textwidth}{!}{
  \begin{tabular}{l|cccccc|cccccc}
    \toprule
    & \multicolumn{6}{c|}{TwiU} & \multicolumn{6}{c}{FliU}  \\ 
    \cline{2-13}
    &In1	&In2	&InT	&InU+1	&InU+T	&NoIn	&In1	&In2	&InT	&InU+1	&InU+T	&NoIn\\
    \midrule
FewUser	&\textbf{71.77} 	&66.19 	&61.72 	&58.53 	&51.36 	&42.74 	&\textbf{56.15} 	&53.83 	&47.49 	&53.14 	&50.55 	&30.43 \\
ClassUser	&60.13 	&\textbf{60.61} 	&46.25 	&50.88 	&50.08 	&24.56 	&43.95 	&\textbf{49.08} 	&26.47 	&37.24 	&32.44 	&3.96 \\
  \bottomrule
  \end{tabular}
  }
\end{table*}

\textbf{How to fuse user features?} 
As detailed in Section \ref{ssec:input}, we utilize three types of user encoders to transform the feature matrix into a single user vector. 
We perform experiments on user encoder designs, with the results presented in Table \ref{tab:user-encoder}, and we select the integration type as In2. 
This is because In1 directly outputs a single vector, thereby obviating the need for fusion, and In2 offers overall suboptimal performance.
Contrary to expectations, time series models, which incorporate temporal information, do not exhibit superior performance in user geolocation tasks. 
Instead, mean pooling, a parameter-free approach, emerges as the most effective method.
\begin{table*}[!htb]
  \caption{Performance of FewUser and ClassUser on TwiU and FliU with various user encoders using the integration type In2. MP and Trans. stand for mean pooling and transformer, respectively.}
  \label{tab:user-encoder}
  \centering
  \begin{tabular}{l|l|cccccccc}
    \toprule
 & Model & MP &	Adapter	&Trans.	&MLP	&LSTM	&BiLSTM	&RNN	&GRU\\
    \midrule
    \multirow{2}{*}{TwiU}	&FewUser	&\textbf{66.19} 	&65.23 	&63.96 	&59.81 	&57.26 	&59.33 	&52.95 	&56.30 \\
    &ClassUser	&\textbf{60.61} 	&60.13 	&58.21 	&59.49 	&50.08 	&36.52 	&57.74 	&55.66 \\
 \hline
    \multirow{2}{*}{FliU}	&FewUser	&53.83 	&56.42 	&\textbf{57.53} 	&50.87 	&48.86 	&53.78 	&49.66 	&47.91 \\
    &ClassUser	&49.08 	&47.12 	&50.66 	&55.47 	&39.67 	&\textbf{56.63} 	&48.55 	&47.81 \\
  \bottomrule
  \end{tabular}
\end{table*}


\subsection{Ablation Studies}
In this section, we present comprehensive ablation studies to examine the influence of textual backbones and geographical prompting on the performance of our model FewUser. 
All reported results are obtained from experiments conducted on the TwiU dataset under an 8-shot setting.

\begin{table}
    \caption{Performance of Fewuser with different textual encoders on TwiU.}
    \centering
    \begin{tabular}{lc|lc}
    \toprule
    Large Model & acc(\%) & Base Model & acc(\%)\\
    \midrule
    SimCSE-BERT-Large & 	\textbf{71.77} &SimCSE-BERT-Base & 	\textbf{64.11} \\
    SimCSE-RoBERTa-Large & 	67.62&    SimCSE-RoBERTa-Base & 	59.81 \\
    -&- & LongFormer-Base & 	58.05 \\
    ALBERT-Large & 	45.61 & ALBERT-Base & 	58.21\\
    BERT-Large & 	5.74 & BERT-Base & 	11.64\\
    RoBERTa-Large & 	6.54 &RoBERTa-Base & 	9.09\\
    BERTweet-Large & 	9.25 &- &-\\
  \bottomrule
    \end{tabular}
    \label{tab:backbone}
\end{table}

\textbf{Textual Backbones.}
We experimented with various textual backbones, and the results are presented in Table \ref{tab:backbone}. The encoders based on the SimCSE series models exhibit the best overall performance, with SimCSE-BERT-Large achieving the highest accuracy of 71.77\%. This superior performance can be attributed to SimCSE's utilization of contrastive learning to enhance sentence embeddings, which synergizes well with our contrastive learning-based model, FewUser.

In comparison, widely used PLMs such as BERT and RoBERTa demonstrate significantly lower performance. 
This might indicate that their training objectives are not as well-aligned with user geolocation tasks as those of SimCSE models.
Surprisingly, BERTweet, a model specifically trained on a large Twitter dataset, only achieves an accuracy of 9.25\%. This may be due to its structural similarity to BERT and its adherence to RoBERTa's pre-training procedures.
Longformer is also involved in comparison to handling the concatenated long input sentences (integrated using In1).
A consistent observation across the models is that the base versions generally underperform compared to their larger counterparts, including those within the SimCSE series.

\begin{table}
    \caption{Performance of FewUser with hard prompts and semi-soft prompts on TwiU.}
    \centering
    \begin{tabular}{lcc}
    \toprule
        Prompt & Hard & Semi-Soft\\
    \midrule
"I'm in [CLASS]." & 	67.62 & 	\textbf{71.77}\\
"A local from [CLASS]." & 	68.74 & 	69.54\\
"[CLASS] in the house!" & 	71.61 & 	70.65\\
"[CLASS] 's own." & 	70.97 & 	68.42\\
"A user resides in [CLASS]." & 	67.15 & 	70.18\\
“Question: where does this user reside in? Answer: [CLASS].” & 	69.06 & 	69.38\\
“Question: which city does this user live in? Answer: [CLASS]." & 	65.55 & 	68.1\\
"[CLASS] " & 	65.87 & 	68.42\\
No Prompt: "[CLASS]" & 	68.9 & 	-\\
  \bottomrule
    \end{tabular}
    \label{tab:hard-semi}
\end{table}

\textbf{Geographical Prompting.}
To investigate the efficacy of geographical prompting, we conduct extensive experiments. 
Due to space constraints, we select only a subset of key results for presentation in Tables \ref{tab:hard-semi} and \ref{tab:soft}. 
Examining the hard prompts, a comparison between the no-prompt baseline and the minimal prompt "[CLASS] " reveals that even the addition of a single space can significantly impact performance, highlighting the method's sensitivity to prompting. 
Furthermore, enhancing lexical diversity, as illustrated by the prompt "A user resides in [CLASS]." and its QA variant, indeed improves performance by 1.31\%. 
Despite requiring manual effort, hard prompts yield commendable results. 
The optimal prompt leads to a performance improvement of 2.71\% over the no-prompt scenario, possibly because it integrates human understanding of the task at hand. 

Semi-soft prompts, however, outperform hard prompts, further enhancing performance by 0.16\%.
This suggests that a balanced method of initializing token vectors, informed by hard prompts yet allowing for flexibility during training, effectively captures the geographical nuances needed for user geolocation tasks.
An intriguing observation is the inconsistency in performance across different templates for hard and semi-soft prompts. 
For instance, the template "I'm in [CLASS]." which excels under semi-soft prompting, performs worse than the no-prompt baseline when used as a hard prompt. 
On the other hand, soft prompts may not fully capitalize on due to their random initialization, exhibit average performance, and lack noteworthy results. 

\begin{table}
    \caption{Performance of FewUser with soft prompts on TwiU.}
    \centering
  \resizebox{1\textwidth}{!}{
    \begin{tabular}{lcccccccccccccc}
    \toprule
       \# Tokens  & 1 & 2 & 3 & 4 & 5 & 6 & 7 & 8 & 9 & 10 & 11 & 12 & 13 & 14\\
    \midrule
acc(\%) &	67.78 & 	60.93 & 	67.78 & 	69.38 & 	66.99 & 	68.42 & 	67.46 & 	\textbf{69.86} & 	65.39 & 	68.58 & 	67.46 & 	64.91 & 	68.9 & 	66.67\\
  \bottomrule
    \end{tabular}
    }
    \label{tab:soft}
\end{table}



\section{Conclusion}
In this work, we presented FewUser, a powerful framework for social user geolocation that outperforms existing models in both zero-shot and various few-shot contexts. 
FewUser integrates a user representation module, leveraging the SimCSE encoder, with a geographical prompting module that utilizes an array of prompts—hard, soft, and semi-soft—to effectively encode locational data.
Our extensive analysis underscores the critical role of input representation in user geolocation. 
Notably, we found that simpler integration methods often lead to superior performance.
Ablation studies on textual backbones and geographical prompting provided further insights into the framework's performance. 
Looking ahead, we are committed to enhancing computational efficiency, possibly through efficient fine-tuning approaches or by advancing zero-shot geolocation capabilities.

\textbf{Ethical Considerations}. 
It is imperative to state that our research utilizes only publicly available user data. We adhere rigorously to ethical standards in research, assuring that no individual's private information is revealed beyond the aggregated data analysis presented herein or what the user has publicly shared.

\bibliographystyle{splncs04}
\bibliography{pkdd2024}

\end{document}